\documentclass[twocolumn,preprintnumbers,aps,showpacs,superscriptaddress]{revtex4-1}

\usepackage{palatino}
\usepackage[latin1]{inputenc}
\usepackage{epsf}
\usepackage{amsmath,amssymb}
\usepackage{latexsym}
\usepackage{calc}
\usepackage{color}
\usepackage{shadow}
\usepackage{epsfig}
\usepackage[pdftex, pdfstartview = {FitH}]{hyperref}

\newcommand{\ben}{\begin{equation}}
\newcommand{\een}{\end{equation}}
\newcommand{\bea}{\begin{eqnarray}}
\newcommand{\eea}{\end{eqnarray}}

\def\ext{_{\rm ext}}

\def\dulR{{\underline{\underline{\bf R}}}}
\def\dulr{{\underline{\underline{\bf r}}}}

\begin{document}
  \title{Exact factorization of the time-dependent electron-nuclear wavefunction} 
  \author{Ali Abedi}
\affiliation{Max-Planck Institut f\"ur Mikrostrukturphysik, Weinberg 2,
D-06120 Halle, Germany}
\affiliation{European Theoretical Spectroscopy Facility (ETSF)}
\author{Neepa T. Maitra} 
\affiliation{Department of Physics and Astronomy, Hunter College and the City University of New York,\\ 695 Park Avenue, New York, New York 10065, USA}
\author{E.K.U. Gross}
\affiliation{Max-Planck Institut f\"ur Mikrostrukturphysik, Weinberg 2,
D-06120 Halle, Germany}
\affiliation{European Theoretical Spectroscopy Facility (ETSF)}

  \date{\today}
  \begin{abstract}
We present an exact decomposition of the complete wavefunction for a
system of nuclei and electrons evolving in a time-dependent external
potential. We derive formally exact equations for the nuclear and electronic
wavefunctions that lead to rigorous definitions of a
time-dependent potential energy surface (TDPES) and a time-dependent geometric
phase. For the $H_2^+$ molecular ion exposed to a laser field, the 
TDPES proves to be a useful interpretive tool to identify different mechanisms of dissociation.
 \end{abstract}
 \maketitle 
 Treating electron-ion correlations in molecules and solids
 in the presence of time-dependent external fields is a major challenge, especially beyond
 the perturbative regime. To make numerical calculations feasible, the description 
 usually involves approximations such as classical dynamics for nuclei with electron-nuclear coupling
 provided by Ehrenfest dynamics or surface-hopping~\cite{TIW96}, or
 even just static nuclei~\cite{CRG09}. Quantum features
 of the nuclear dynamics (e.g., zero-point energies, tunneling, and
 interference) are included approximately in some methods~\cite{HBNS06,Martinez}, while numerically exact solutions of the
 time-dependent Schr\"odinger equation (TDSE) for the coupled system of
 electrons and nuclei have been given for very small systems like
 $H_2^+$~\cite{Bandrauk}. Clearly, the full electron-nuclear wavefunction contains the complete information on the system, but it lacks the intuitive
 picture that potential energy surfaces (PES) can provide. To this end, approximate TDPES were introduced by Kono~\cite{KSTK04} as
 instantaneous eigenvalues of the electronic Hamiltonian, and proved
 extremely useful in interpreting system-field phenomena.
 The concept of a TDPES arises in a different way in Cederbaum's recent work, where
 the Born-Oppenheimer (BO) approximation is generalized to
 the time-dependent case~\cite{C08}.

In the present Letter we provide a rigorous separation
of electronic and nuclear motion by introducing an {\it exact}
factorization of the full electron-nuclear
wavefunction. The factorization is a natural extension of the work of Hunter~\cite{Hunter}, in which an exact decomposition was developed for the static problem. It leads to an exact definition of the TDPES as
well as a Berry vector potential. Berry-Pancharatnam phases~\cite{Berry} are
usually interpreted as arising from an approximate decoupling of a system from
``the rest of the world'', thereby making the system Hamiltonian dependent on some ``environmental'' parameters. For example, in the static BO approximation, 
the electronic Hamiltonian depends parametrically on the nuclear positions; i.e., the stationary electronic Schr\"odinger equation is solved for each fixed nuclear configuration $\dulR$, yielding $\dulR$-dependent eigenvalues (the BO PES) and eigenfunctions (the BO wavefunctions). If the total molecular wavefunction is approximated by a single product of a BO wavefunction and a nuclear wavefunction, the equation of motion of the latter contains a Berry-type vector potential. One may ask: is the appearance of Berry phases a consequence of the BO approximation or does it survive in the exact treatment?
In this Letter we demonstrate that even in the exact treatment of the electron-nuclear system a Berry connection
appears and we prove a new relation between this connection and the nuclear velocity field.
For a numerically exactly solvable system we calculate the exact TDPES, demonstrate their
interpretive power, and compare with approximate treatments.
Throughout this paper we use atomic units and the electronic and nuclear coordinates are collectively denoted by $\dulr$, $\dulR$.  
The Hamiltonian for a system of interacting electrons and nuclei, evolving under a time-dependent external potential, may be written as
 \begin{equation}
    \hat{H} = \hat{H}_{BO} + \hat{V}\ext^e(\dulr,t)+\hat{T}_n(\dulR) + \hat{V}\ext^n(\dulR,t)
  \end{equation}
 where $\hat H_{BO}$ is the traditional BO electronic Hamiltonian,
\ben
\hat{H}_{BO} = \hat{T}_e(\dulr) + \hat{W}_{ee}(\dulr) + \hat{V}_{en}(\dulr,\dulR) + \hat{W}_{nn}(\dulR).  
\label{eq:tradBO}
\een
Here $\hat T_n$($\hat T_e$) is the nuclear(electronic) kinetic energy operator,
$\hat{W}_{nn}$ ($\hat{W}_{ee}i$) is the nuclear-nuclear (electron-electron) interaction, and 
 $\hat{V}\ext^n(\dulR,t)$ and $\hat{V}\ext^e(\dulr,t)$ are time-dependent external potentials acting on the nuclei and electrons, respectively.
The complete electron-nuclear wavefunction satisfies the TDSE:
\ben
\hat{H} \Psi(\dulr,\dulR,t)= i\partial_t\Psi(\dulr,\dulR,t).
\label{eq:tdse}
\een
The central statement of this Letter is the following:

{\it Theorem I.} (a) The {\it exact} solution of Eq.~(\ref{eq:tdse}) can be written as a single product 
\begin{equation}
  \Psi(\dulr,\dulR,t)=\Phi_{\dulR}(\dulr,t)\chi(\dulR,t)
\label{eq:Psi}
\end{equation}
where $\Phi_{\dulR}(\dulr,t)$ satisfies the normalization condition, 
\ben
\int d\dulr\vert\Phi_{\dulR}(\dulr,t)\vert^2=1 \;,
\label{eq:partialnorm}
\end{equation}
for any fixed nuclear configuration, $\dulR$, at any time $t$. 

(b) The wavefunctions $\Phi_{\dulR}(\dulr,t)$ and $\chi(\dulR,t)$ satisfy: 

\ben
  \label{eq:exact_el_td}       
  \Bigl(\hat{H}_{el}(\dulr,\dulR,t)-\epsilon(\dulR,t)\Bigr)\Phi_{\dulR}(\dulr,t)\\=i\partial_t \Phi_{\dulR}(\dulr,t),
\een
\ben
  \label{eq:exact_n_td}
  \begin{split}
  \Bigl(\sum_{\nu=1}^{N_n}&\frac{1}{2M_\nu}(-i\nabla_\nu+{\bf A}_\nu(\dulR,t))^2 \\
  &+\hat{V}_{ext}^n(\dulR,t) + \epsilon(\dulR,t)\Bigr)\chi(\dulR,t)=i\partial_t \chi(\dulR,t),
\end{split}             
\een
where the electronic Hamiltonian is
\ben
\label{eq:e_ham_td}
\begin{split}
\hat{H}_{el}(\dulr,\dulR,t) = &\hat{H}_{BO}+\hat{V}\ext^e(\dulr,t) +\sum_{\nu=1}^{N_n}\frac{1}{M_\nu} \\ 
&\times\Big[\frac{(-i\nabla_\nu-{\bf A}_\nu(\dulR,t))^2}{2} + \Big(\frac{-i\nabla_\nu \chi}{\chi}\\
&+{\bf A}_\nu(\dulR,t)\Big)\left(-i\nabla_\nu-{\bf A}_\nu(\dulR,t)\right)\Big].
\end{split}
\een

Here the scalar and vector potential terms are

\bea
  \label{eq:exact_eps_td}
  &&\epsilon(\dulR,t) = \left\langle\Phi_{\dulR}(t) \right\vert\hat{H}_{el}((\dulr,\dulR,t) - i \partial_t\left\vert \Phi_{\dulR}(t)\right\rangle_\dulr \\
  \label{eq:exact_BP_td}
  &&{\bf A}_\nu(\dulR,t)=\left\langle\Phi_{\dulR}(t)\right\vert\left.-i\nabla_\nu\Phi_\dulR(t)\right\rangle_\dulr
\eea
where $\langle ..|..|..\rangle_\dulr$ denotes an inner product over all electronic variables only. 

{\it Proof.} Part (a): We must show that the exact solution $\Psi(\dulr,\dulR,t)$ of the full
TDSE~(\ref{eq:tdse}) can be factorized as in Eqs.~(\ref{eq:Psi})-(\ref{eq:partialnorm}). To show this, choose at each point in time
$\chi(\dulR,t) =  e^{iS(\dulR,t)}\sqrt{\int d\dulr\vert\Psi(\dulr,\dulR,t)\vert^2}$ and
$\Phi_{\dulR}(\dulr,t) = \Psi(\dulr,\dulR,t) / \chi(\dulR,t)$, where $S(\dulR,t)$ is real. 
The normalization condition~(\ref{eq:partialnorm}) then follows immediately.

Part (b): To derive Eqs.~(\ref{eq:exact_el_td})-(\ref{eq:exact_BP_td}), we apply Frenkel's stationary action principle, $\delta\int_{t_0}^{t_1} dt\langle\Psi\vert \hat{H} - i\partial_t\vert \Psi\rangle = 0\;,
$ to the wavefunction~(\ref{eq:Psi}). We require the action to be stationary with respect to variations in $\Phi_\dulR(\dulr, t)$ and $\chi(\dulR, t)$, 
subject to the condition~(\ref{eq:partialnorm}). This then leads, after some algebra, to Eqs.~(\ref{eq:exact_el_td})-(\ref{eq:exact_BP_td}).
 Hence, the product wavefunction~(\ref{eq:Psi}) is a stationary point of the action functional, but we still have to prove that this stationary point
 corresponds to an {\it exact} solution of the TDSE. By evaluating 
 $i\partial_t(\Phi_{\dulR}(\dulr,t)\chi(\dulR,t))$ 
 and inserting Eqs.~(\ref{eq:exact_el_td})-(\ref{eq:exact_n_td}), we verify that the full TDSE~(3) is satisfied.
 
{\it Theorem II.}
(a) Eqs.~(\ref{eq:exact_el_td})-(\ref{eq:e_ham_td}) are form-invariant under the following gaugelike transformation
\bea
\nonumber  
&&\Phi_{\dulR}(\dulr,t)\rightarrow\tilde{\Phi}_{\dulR}(\dulr,t)=\exp(i\theta(\dulR,t))\Phi_{\dulR}(\dulr,t)\\
&&\chi(\dulR,t)\rightarrow\tilde{\chi}(\dulR,t)=\exp(-i\theta(\dulR,t))\chi(\dulR,t)
\label{eq:phiT}
\eea

\bea
\label{eq:GT}
\nonumber
       &&\mathbf{A}_\nu(\dulR,t)\rightarrow\tilde{\mathbf{A}}_\nu(\dulR,t)=\mathbf{A}_\nu(\dulR,t)+\nabla_{\nu}\theta(\dulR,t) \\
       &&\epsilon(\dulR,t)\rightarrow\tilde{\epsilon}(\dulR,t)=\epsilon(\dulR,t) + \partial_t\theta(\dulR,t) 
\eea

(b) The wavefunctions $\Phi_{\dulR}(\dulr,t)$ and $\chi(\dulR,t)$ yielding a given solution, $\Psi(\dulr,\dulR,t)$, of Eq.~(\ref{eq:tdse}) are unique up to within the ($\dulR,t$)-dependent phase transformation~(\ref{eq:phiT}).

{\it Proof.} The form invariance of Eqs.~(\ref{eq:exact_el_td})-(\ref{eq:e_ham_td}) is easily verified by inserting Eqs.~(\ref{eq:phiT})-(\ref{eq:GT}) into 
Eqs.~(\ref{eq:exact_el_td})-(\ref{eq:e_ham_td}) which proves part (a). To prove part (b), assume the exact wavefunction
can be represented by two different products: 
$\Psi(\dulr,\dulR,t)=\Phi_{\dulR}(\dulr,t)\chi(\dulR,t)=\tilde{\Phi}_{\dulR}(\dulr,t)\tilde{\chi}(\dulR,t)$. 
Defining 
$g(\dulR,t):=\chi(\dulR,t) / \tilde{\chi}(\dulR,t)$, 
then
$\vert \tilde{\Phi}_{\dulR}(\dulr,t) \vert^2 = \vert g(\dulR,t) \vert^2 \vert \Phi_{\dulR}(\dulr,t) \vert^2$.
Integrating this over $\dulr$ and using Eq.~(\ref{eq:partialnorm}), we get $\vert g(\dulR,t) \vert^2 = 1$ implying $g(\dulR,t) = e^{i\theta(\dulR,t)}$ and hence the desired result
$\tilde{\Phi}_{\dulR}(\dulr,t)=e^{i\theta(\dulR,t)}\Phi_{\dulR}(\dulr,t)$.

The wavefunctions $\Phi_{\dulR}(\dulr,t)$ and $\chi(\dulR,t)$ have a clear-cut physical meaning:
$\vert\chi(\dulR,t)\vert^2 =\int \vert\Psi(\dulr,\dulR,t)\vert^2 d\dulr$ is the probability density of finding the nuclear configuration $\dulR$ at time $t$, and
$\vert\Phi_{\dulR}(\dulr,t)\vert^2 = \vert\Psi(\dulr,\dulR,t)\vert^2 / \vert \chi(\dulR,t)\vert^2$ is the conditional probability of finding the electrons at $\dulr$, given that the nuclear configuration is $\dulR$. At locations where $\vert\chi(\dulR,t)\vert^2$ approaches zero
the TDPES may show peaks, in close analogy to the ``quantum potential'' in the Bohmian formulation of quantum mechanics.\\
Eqs.~(\ref{eq:exact_el_td})-(\ref{eq:exact_BP_td}) determine the {\it exact} time-dependent molecular wavefunction, given an initial state. As written, the nuclear equation 
is particularly appealing as a Schr\"odinger equation with both scalar and vector-potential coupling terms contributing effective forces on the nuclei 
including any geometric phase effects. We call $\epsilon(\dulR,t)$ and ${\bf A}(\dulR,t)$ the exact TDPES and time-dependent Berry connection,
respectively. These two quantities mediate the coupling between the
nuclear and the electronic degrees of freedom in a formally exact way. Eqs.~(\ref{eq:exact_el_td})-(\ref{eq:exact_BP_td}) demonstrate that a Berry connection indeed appears in the exact treatment. But does it produce a real effect or
can it be gauged away by a suitable choice of $\theta(\dulR,t)$ in Eqs.~(\ref{eq:phiT})-(\ref{eq:GT})? To shed some light on this question, we now prove an alternate expression
for the vector potential. Inserting $\Phi_\dulR = \Psi/\chi$ into Eq.~(\ref{eq:exact_BP_td}), and evaluating the nuclear gradient on this quotient,
reveals that it is the difference of paramagnetic nuclear velocity fields derived from the full and nuclear wavefunctions:
\ben
  \label{eq:exact_vect}
  {\bf A}_\nu(\dulR,t)=\frac{Im \left\langle\Psi(t)\right\vert\left.\nabla_\nu\Psi(t)\right\rangle_\dulr}{\vert \chi(\dulR,t)\vert^{2}} - \frac{Im (\chi^*\nabla_\nu\chi)}{\vert \chi(\dulR,t)\vert^{2}}. 
\een

This equation is interesting in several respects. First, writing $\chi(\dulR,t) =  e^{iS(\dulR,t)}\vert\chi(\dulR,t)\vert$, the last term
on the right-hand-side of Eq.~(\ref{eq:exact_vect}) can be represented as $\nabla_{\nu} S(\dulR,t)$, so it can be gauged away. Consequently, any true Berry connection
(that cannot be gauged away) must come from the first term. If the exact $\Psi(t)$
is real-valued (e.g. for a non-current-carrying ground state) then the first term on the right-hand-side of Eq.~(\ref{eq:exact_vect}) vanishes and hence the exact Berry connection
vanishes. Second, since $Im \left\langle\Psi(t)\right\vert\left.\nabla_\nu\Psi(t)\right\rangle_\dulr$  is the
true nuclear (many-body) current density,  Eq.~(\ref{eq:exact_vect}) implies that the gauge-invariant current density, $Im (\chi^*\nabla_\nu\chi)+\vert \chi\vert^{2}{\bf A}_\nu$, 
 that follows from Eq.~(\ref{eq:exact_n_td}) does indeed reproduce the exact nuclear current density~\cite{Manz}. Hence, the solution $\chi(\dulR,t)$ of Eq.~(\ref{eq:exact_el_td}) is, in every respect, the proper nuclear many-body wavefunction: Its absolute-value squared gives the exact nuclear ($N$-body) density while its phase yields the correct 
nuclear ($N$-body) 
current density.

 In the following, we first discuss some limiting cases of the exact Eqs.~(\ref{eq:exact_el_td})-(\ref{eq:exact_BP_td}). Fixing the gauge via 
$\left\langle\Phi_{\dulR}(t) \vert \partial_t \Phi_{\dulR}(t)\right\rangle_\dulr \equiv 0$, the electronic equation reads
\begin{equation}
 \label{eq:eleceqn_infmass}
 \hat{H}_{el}(\dulr,\dulR,t)\phi_{\dulR}(\dulr,t)=i\partial_t\phi_{\dulR}(\dulr,t),
\end{equation}
with $ \phi_\dulR = e^{-i\int^t\epsilon(\dulR,\tau) d\tau} \Phi_\dulR$ while the nuclear equation retains its form Eq.~(\ref{eq:exact_n_td}) with
$\epsilon(\dulR,t) = \left\langle\Phi_{\dulR}(t) \right\vert\hat{H}_{el}(\dulr,\dulR,t)\left\vert \Phi_{\dulR}(t)\right\rangle_\dulr $.
 Note that the electronic Eq.~(\ref{eq:eleceqn_infmass}) and the nuclear Eq.~(\ref{eq:exact_n_td}) have to be propagated simultaneously because the Hamiltonian~(\ref{eq:e_ham_td})
 depends on $\chi(\dulR,t)$. Taking the large-nuclear-mass limit of Eq.~(\ref{eq:e_ham_td}), the electronic
 Hamiltonian reduces to ${\small \hat{H}_{el}\xrightarrow[M\to\infty]{}\hat{H}_{BO}+\hat{V}_{ext}^e}$, i.e., the dependence of $H_{el}$ on $\chi$ drops out and
 the electronic Eq.~(\ref{eq:eleceqn_infmass}) depends on the nuclear configuration $\dulR$ only parametrically. So, in this limit, 
 Eq.~(\ref{eq:eleceqn_infmass}) is propagated in time for each fixed nuclear configuration $\dulR$, which is precisely Cederbaum's time-dependent
 generalization of the BO approximation~\cite{C08}. Hence, the full Eqs.~(\ref{eq:exact_el_td})-(\ref{eq:exact_BP_td}) can be viewed as an ``exactification'' of the intuitively appealing procedure of 
 Ref.~\cite{C08}. If, furthermore, we treat the nuclei classically, i.e., use the Hamiltonian~(\ref{eq:exact_n_td}) to generate classical equations of
 motion for the nuclei, we obtain
\begin{equation}
  \label{eq:nuc_climit}
  M\ddot\dulR_{\nu} = {\bf E_\nu} + \dot\dulR_{\nu} \times {\bf B_\nu}
\end{equation} 
where the electric and magnetic ''Berry fields'' are given by
${\bf E_{\nu}} = \nabla_{\nu} \epsilon(\dulR,t)  - \frac{\partial \bf A_{\nu}}{\partial t}$ and ${\bf B_\nu} = \nabla_{\nu} \times {\bf A_\nu}$.
The additional magnetic field was also found, in the appropriate limit, in an exact path-integral approach to the coupled dynamics~\cite{K07}, and also in other work~\cite{ZW06}.
Being strictly equivalent to the TDSE, the electronic and nuclear Eqs.~(\ref{eq:exact_el_td})-(\ref{eq:exact_n_td}) provide a rigorous starting point suitable for making 
systematic semiclassical approximations~\cite{Ciccotti,Burke} beyond the purely classical limit of Eq.~(\ref{eq:nuc_climit}).

\begin{figure}[h]
\centering
\includegraphics[width=1\columnwidth]{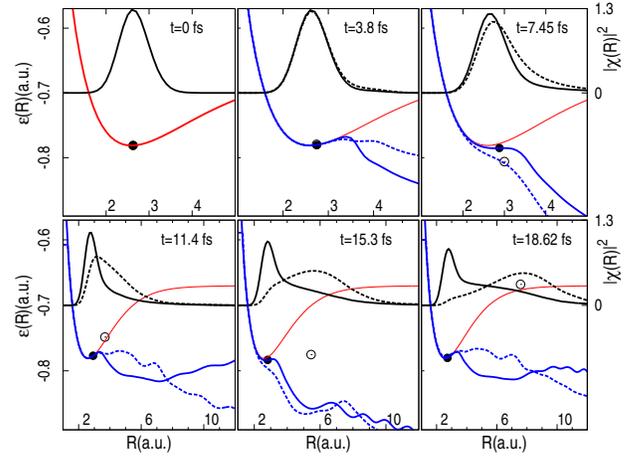}
\caption{Snapshots of the TDPES (blue lines) and nuclear density (black) at times indicated, for the H$_2^+$ molecule subject to the laser-field (see text),  $I_1 = 10^{14}$W/cm$^2$ (dashed line) and $I_2 = 2.5 \times 10^{13}$W/cm$^2$ (solid line). The circles indicate the position and energy of the classical particle in the exact-Ehrenfest calculation ($I_1$: open, $I_2$: solid). For reference, the ground-state BO surface is shown as the thin red line.}
\label{fig:tdpes1}
\end{figure}

We now return to the exact formulation to investigate the TDPES for a numerically exactly solvable model: the $H_2^+$ molecular
ion subject to a linearly polarized laser field. By restricting the motion of the nuclei and the electron to the direction of the polarization axis of the laser field
, the problem can be modelled with a 1D Hamiltonian featuring ``soft-Coulomb'' interactions~\cite{LKG}:
\bea
  \label{eq:H2+}
  \begin{split}
    \hat{H}(t) = & - \frac{1}{M}\frac{\partial^2}{\partial R^2} - \frac{1}{2\mu_e}\frac{\partial^2}{\partial z^2}+ \frac{1}{\sqrt{0.03+R^2}} + \hat{V}_l(z,t)\\
    &- \frac{1}{\sqrt{1+(z-R/2)^2}} - \frac{1}{\sqrt{1+(z+R/2)^2}} \\     
  \end{split}
\eea
where $R$ and $z$ are the internuclear distance and the electronic coordinate as measured from the nuclear center-of-mass,
respectively, and the electronic reduced mass is given by $\mu_e=(2M)/(2M+1)$, $M$ being the proton mass. The laser field 
is represented by $\hat{V}_l(z,t) = q_ezE(t)$ where
$E(t)$ denotes the electric field amplitude and the reduced charge $q_e =(2M+2)/(2M+1)$. We consider
a 
$\lambda = 228$ nm laser field,  represented by $E(t) = E_0f(t)\sin(\omega t)$,
for two peak intensities, $I_1 = |E_0|^2 = 10^{14} W/$cm$^2$ and 
$I_2 = |E_0|^2 = 2.5 \times 10^{13} W/$cm$^2$. The envelope function $f(t)$ is chosen such that the field is linearly ramped from zero
to its maximum strength at $t=7.6$ fs and  thereafter held constant.

\begin{figure}[h]
\centering
\includegraphics[width=1\columnwidth]{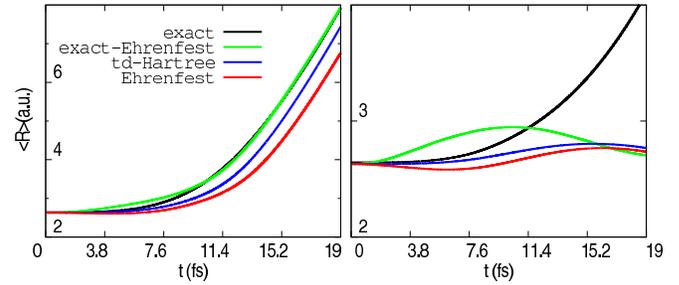}
\caption{The internuclear separation $\langle R\rangle(t)$ for the same intensities as in Fig.1. Left panel : $I_1$. Right panel : $I_2$.}
\label{fig:R1}
\end{figure}

 Starting from the exact ground-state as initial condition, we propagate the TDSE numerically to obtain 
 the full molecular wavefunction $\Psi(z,R,t)$.
As there is only one nuclear degree of freedom (after separating off the center-of-mass motion), we can fix the gauge
in Eqs.~(\ref{eq:phiT})-(\ref{eq:GT}) such that the vector potential~(\ref{eq:exact_vect}) is always zero. 
From the computed exact time-dependent molecular wavefunction we compute the TDPES's; these,
along with the corresponding nuclear density, $\vert\chi(R,t)\vert^2$, are
plotted in Fig.~\ref{fig:tdpes1} at six snapshots of time.
 The initial TDPES lies practically on top of the ground-state BO
 surface, which is plotted in all the snapshots for comparison. 
Fig.~\ref{fig:R1} shows the exact internuclear
distance $\left\langle \Psi(t)\right\vert \hat{R} \left.\vert
\Psi(t)\right\rangle$, along with the results from three approximate
methods: (i) the usual Ehrenfest approximation (i.e. Eq.~\ref{eq:nuc_climit}),
(ii) the ``exact-Ehrenfest'' approximation, which substitutes the exact TDPES for the Ehrenfest potential in the usual Ehrenfest approach
and, (iii) an uncorrelated approach, the time-dependent Hartree (self-consistent field) approximation, $\Psi_H(\dulr,\dulR,t)=\phi(\dulr,t)\chi(\dulR,t)$, where the electronic part
does not depend on $\dulR$ at all. Fig.~\ref{fig:R1} shows that for the intensity $I_1$, 
all methods yield dissociation, while for the weaker $I_2$, only the exact does. We now discuss how the TDPES contains the signature of this behavior.  
Note that the laser-field does not couple directly to the nuclear relative coordinate $R$, but only indirectly via the TDPES.

$I_1 = |E_0|^2 = 10^{14} W/$cm$^2$: The dissociation of the molecule is dramatically reflected in the exact TDPES, whose well flattens out, causing the nuclear density to
 spill to larger separations. Importantly, the tail of the TDPES alternately falls sharply and returns in correspondence
 with the field, letting the density out; the TDPES is the only potential acting on the nuclear system and transfers energy from the accelerated electron to the nuclei.
The expectation value of the internuclear distance in Fig.~\ref{fig:R1}, demonstrates that 
among all the approximate calculations employed here, the exact-Ehrenfest is most accurate. Surprisingly, it even does better than TD-Hartree
which treats the protons quantum mechanically, thus showing the importance of electron-nuclear correlation.

$I_2 = |E_0|^2 = 2.5 \times 10^{13} W$/cm$^2$ : From Fig.~\ref{fig:R1}, the exact calculation leads to dissociation, while none of the approximations do, in contrast to the
 previous case. The TDPES of Fig.~\ref{fig:tdpes1}, suggests that tunneling is the leading mechanism for the dissociation: a well remains at all times that traps a 
classical particle, which would oscillate inside it, as indeed reflected in Fig.~\ref{fig:R1}. (See also the solid circles in Fig.~\ref{fig:tdpes1}). Although the tail has  
similar oscillations as for $I_1$, this does not lead to dissociation of classical nuclei due to the barrier; the TDPES in this case transfers the 
field energy to the nuclei via tunneling. Although the exact-Ehrenfest has a larger amplitude of oscillation than the others, it ultimately cannot
tunnel through the barrier.  

This example demonstrates how studying the TDPES reveals the mechanism
of dissociation. Because the TDPES includes the electron-nuclear
correlation exactly, we believe the exact-Ehrenfest dynamics is the
best one could do within a classical treatment of the nuclei. There is a need to go beyond classical dynamics
when the dissociation proceeds mainly via tunneling.

In conclusion, we have presented a rigorous factorization of the complete molecular wavefunction into an electronic contribution, $\Phi_{\dulR}(\dulr,t)$, and a nuclear part, $\chi(\dulR,t)$. 
The exact nuclear $N$-body density is $\vert\chi(\dulR,t)\vert^2$ while $\vert\Phi_{\dulR}(\dulr,t)\vert^2$ 
represents the conditional probability of finding the electrons at $\dulr$, given the nuclear configuration $\dulR$. 
Their exact equations 
of motion are deduced. Via these equations, the time-dependent potential energy 
surface (\ref{eq:exact_eps_td}) and the time-dependent Berry connection (\ref{eq:exact_BP_td}) are defined as rigorous concepts. We demonstrated with numerical examples that
the TDPES is a powerful tool to analyze and interpret different types of dissociation processes (direct vs tunneling). The exact splitting of electronic and nuclear degrees 
of freedom presented by Eqs.~(\ref{eq:exact_el_td})-(\ref{eq:exact_BP_td}) lends itself as a rigorous starting point for making approximations, especially for the systematic
development of semiclassical approximations. As a first step we have shown how the Ehrenfest equations with Berry potential emerge from treating the nuclei classically in the
large-nuclear-mass limit. 

This work was supported by the European Community through the e-I3 ETSF project (INFRA-2007-211956), the National Science 
Foundation grant CHE-0647913 and a Research Corporation Cottrell Scholar Award. We thank Angelica Zacarias for comments.

\end{document}